\input epsf
\documentclass[11pt]{article}

\usepackage{amsmath}
\usepackage{amsfonts}
\usepackage{amssymb}
\usepackage{paper2e}

\newcommand{\cb}{\bar{c}}
\newcommand{\bra}{\langle}
\newcommand{\ket}{\rangle}
\newcommand{\cH}{{\cal H}}
\newcommand{\cd}{{c^\dagger}}

\newcommand{\al}{\alpha}
\newcommand{\be}{\beta}
\newcommand{\g}{\raisebox{.3mm}{$\gamma$}}
\newcommand{\M}{{_{(M)}}}
\newcommand{\vt}{\widetilde{v}\, }
\newcommand{\alphas}{{\al_1 \al_2 \dots \al_N}}
\newcommand{\betas}{{\be_1 \be_2 \dots \be_N}}
\newcommand{\ct}{\widetilde{c}}
\newcommand{\ctd}{{\widetilde{c}^\dagger}}
\newcommand{\mun}{{m_1}}
\newcommand{\md}{{m_2}}
\newcommand{\Si}{{\, \Sigma}}

\def\beq{\begin{equation}}
\def\eeq{\end{equation}}

\begin{document}
\begin{titlepage}

\title{Glimmers of a pre-geometric perspective}

\author{Federico Piazza}

\address{Institute of Cosmology and Gravitation, University of Portsmouth,\\ 
Portsmouth PO1 2EG, United Kingdom}

\begin{abstract}
Spacetime measurements and gravitational experiments are made by using  
objects, matter fields or particles and their mutual relationships. As a consequence,
any operationally meaningful assertion about spacetime is in fact an assertion about the
degrees of freedom of the matter (\emph{i.e.} non gravitational) fields; those, say 
for definiteness, of the Standard Model of particle physics. As for any quantum theory, 
the dynamics of the matter fields can be described in terms of a unitary evolution
of a state vector in a Hilbert space. 
By writing the Hilbert space as a generic tensor product of ``subsystems'' we 
analyse the evolution of a state vector
on an information theoretical basis and attempt to recover the usual spacetime relations
from the information exchanges between these subsystems. 
We consider generic interacting second quantized models with a finite number of 
fermionic degrees of freedom and characterize on physical grounds the tensor product 
structure associated with the class of ``localized systems'' and therefore with 
"position". We find that in the case of free theories no spacetime relation is 
operationally definable. On the contrary, by applying the same procedure to the 
simple interacting model of a one-dimensional 
Heisenberg spin chain we recover the tensor product structure usually associated 
with ``position''.
Finally, we discuss the possible role of gravity in this framework.

\end{abstract}

\end{titlepage}

\section{INTRODUCTION}

Gravity and geometry are so strongly connected at the classical level 
that quantum gravity is generally supposed to be, 
to some extent, a pre-geometric theory \cite{pregeo} i.e. a theory whose 
degrees of freedom are not associated with the points of 
spacetime and where spacetime continuum emerges
in some appropriate coarse-grained limit.
All the questions related to the fine -- Planck scale? -- 
structure of spacetime are therefore traditionally considered as jobs for 
the candidate theories of quantum gravity and associated to substantially
``new''  physics. The breakthrough that such theories 
call for, however, is huge and the directions to be taken while attempting a formulation
highly arbitrary, due to the lack of indications from experiments.

Loop quantum gravity, which takes the pre-geometric issue seriously, is a 
non-perturbative quantization of general relativity (GR). One may object, however, that 
many different theories have GR as a low energy limit and/or that 
at very high scales it may be reasonable to expect a unification between
gravity and the other interactions.
String theory, so rich in powerful insights in its unified framework, still lacks of a totally non-perturbative, pre-geometric formulation: all we can handle at the moment is the dynamics of strings and branes living on a somehow pre-defined spacetime. Again, exploring in the direction of the ``true (M-)theory'' appears to require an extraordinary effort.

The purpose of this paper is to show how a collection of events and some of their mutual metric
properties can emerge from elementary quantum theories formulated without any allusion to a pre-existing space. We try to give 
a constructive procedure for interpreting \emph{a posteriori} as ``spacetime based" a theory that, \emph{a priori}, is not. 
In other words, at this stage, we do not invoke new physics and try to get some pre-geometric insight from theories that -- at least in their usual spacetime-formulations -- are already known.

The spirit aims to be as close as possible to those starting steps in the construction of special relativity that lead to Lorentz transformations. Each inertial observer was given a recipe for the synchronization of its clocks and an operational definition of simultaneity. 
No really new physics was introduced at that stage: the constancy of the velocity of light was already an established -- although quite puzzling -- result. The essential ingredient, rather, was to recognize the physical procedure through which different inertial observers assign the coordinates to the physical events\footnote{In Einstein's words: "Like every other electrodynamics, the theory to be developed is based on the 
kinematics of the rigid body, since assertions of each and any theory concern 
the relations between rigid bodies (coordinate systems), clocks, and 
electromagnetic processes. Insufficient regard for this circumstance is at
the root of the difficulties with which the electrodynamics of moving bodies
must presently grapple''. And even also ``If, for example, I say that --the 
train arrives here at 7 o'clock, that means, more or less, -- the
pointing of the small hand of my clock to 7 and the arrival of the train
are simultaneous events''\cite{eins}.}. 
For the problem at hand we want to attempt an analogous bottom-up approach; leave gravity alone for a while and -- to begin with -- try to ask about spacetime in those theories with which spacetime is actually probed. We refer basically to the Standard Model of Particle Physics or to some unifying and/or supersymmetric extension of it: those are the tools that we actually use in each and every measurement of distance or time interval.

Unfortunately, any attempt to describe the process by which some observer assigns a position to an object eventually
faces the delicate measurement problem of quantum mechanics. In fact -- as we are assuming here that physical theories are quantum mechanical -- every measurement, every acquisition of information about a system by another system is ultimately a quantum measurement. A particle, for instance, does not mark any point in space at any time until it is measured, say, by a screen or a photographic plate. It is the \emph{detection} of a particle -- rather than the particle itself -- which most likely defines a point in spacetime \cite{haag}.  
The Copenhagen interpretation of the measurement process, while perfectly fitting the needs of any experiment or laboratory, does not provide a satisfactory theoretical framework for our purposes:  the act of measuring is there described through a non-unitary evolution and physical systems are divided, without any definite prescription, into ``observers" and ``observed", ``classical" and ``quantum". As a consequence, spacetime relations may not be operationally defined without involving some ``external" observer with its own classical dynamics. 

Without any claim of being exhaustive in such a debated issue nor pretending to solve all the unease that quantum mechanical ``paradoxes" may possibly cause,
we sketch in Sec. \ref{relational} the interpretative-scheme/working-hypotheses to be followed in the rest of the paper. We basically assume that any
measurement is described by a unitary evolution in the tensor product of the Hilbert spaces of the ``observer" and the ``observed".
Correlations play a central role in Everett's view of quantum mechanics. In his seminal 
dissertation \cite{everett}, the relation between quantum correlations and mutual information is deeply exploited and measurements are consistently described as appropriate unitary evolutions that increase the degree of correlation between two subsystems: the ``measured'' and the ``measuring''. By taking Everett's view one can try to re-interpret the evolution of system
as measurements actually going on between the different \emph{subsystems}. 
No fundamental distinction is made between a  measuring apparatus and a particle nor between observing human beings and quantum systems. It is very 
compelling that \emph{local observables} may be eventually picked out within such an abstract scheme.
This is therefore a promising starting point in order to re-interpret as ``spacetime based" the dynamics of simple quantum systems without invoking the presence of complex measuring apparata or, what would be even more problematic, that of a conscious mind.

 As already mentioned, our ``zero-assumption"  is that the physical theories that we use to define spacetime relations are quantum mechanical and that all their physical content can be encoded in the evolution of a state vector $|\Psi(t)\ket$ (the ``state vector of the Universe") in a Hilbert space $\cH_{\rm Universe}$. We can consider quite arbitrary \cite{paolo1} tensor decompositions of $\cH_{\rm Universe}$ into subsystems, 
``parties'' or ``parts'', 
 \begin{equation} \label{intro:1}
\cH_{\rm Universe} \ = \ \cH_{\rm system\ 1} \otimes \cH_{\rm system\ 2} \otimes \dots \ ,
\end{equation}
without any reference to their localization or geometrical relations. 
 The first and most elementary type of spacetime relation that one may attempt to define between these subsystems 
is that of \emph{mutual spacetime coincidence} \emph{i.e.} ``being in the same place at the same time". The quantum-measurement discussion of Sec. \ref{relational} has been given with the aim of giving such a definition some operational meaning. In any experimenter's experience
``having been coincident" with a field or with some of its degrees of freedom has
the precise and definite meaning of ``having detected a particle". The assumptions of 
Sec. \ref{relational} come now into play because, according to them, any such 
``measurement experience" can be described  
within the usual formalism of quantum mechanics and, more importantly, can be recognized
as ``happening'' between some of the parties \eqref{intro:1} once the evolution of $|\Psi(t)\ket$ is known. 
Those readers not interested in the ``quantum measurement'' discussion find the basic assumptions
of Sec. \ref{relational} summarized at the points (I) and (II) of Sec. \ref{qrv} and can directly skip to Sec. \ref{lines} which is the core of the paper.
In Sec. \ref{cont} we define coincidence in terms of quantum entanglement. 

An arbitrary partition of a set of degrees of freedom into ``subsystems'', 
although perfectly legitimate, is generally going to give a highly non-localized description 
of physical events.  In eq. \eqref{intro:1}, for instance, \emph{system}1 may in principle 
account for
completely non-localized degrees of freedom such as, say, the spin of an electron in the Andromeda galaxy
and the position of a cosmic ray entering the Earth atmosphere.
In section \ref{lines:local} 
we attempt to characterize on physical grounds the class of subsystems to which we usually 
attribute the property of \emph{being localized}. The corresponding tensor product structure (TPS), we argue, is the one with respect to which 
the tendency to create coincidence relations is minimal. 

Admittedly, in this pre-geometric scheme, we need an ``external'' time parameter $t$ that 
governs the evolution of the state vectors and, on probabilistic grounds, the fluxes of information between the subsystems.  Nevertheless, the time as perceived 
by each subsystem/observer and thus the overall spacetime picture that emerges from the 
\emph{coincidence relations} between \emph{localized parties} have, in principle,
nothing to do with the parameter $t$. The difference between the external ``inaccessible'' time $t$ and
the time as perceived by the observers is explored in section \ref{time}. 
Here we also note that, in the case of free field theories 
(see Sec. \ref{model:evolution} for details), despite the ubiquitous presence of $t$, 
a spacetime is not even operationally definable.

In Sec. \ref{model} we apply the ideas of Sec. \ref{lines} and consider some definite type of
Hamiltonian $H$ governing the dynamics of $\cH_{\rm Universe}$.
We borrow from the second-quantized formalism of field theories
a toy model with $N$ interacting fermionic degrees of freedom. The toy-Universe is thus 
described by a $2^N$-dimensional Hilbert space and the coincidence relations between 
subsystems can now be quantitatively studied. 
We give, in such a framework, some mathematical instruments for defining tensor 
product structures and for measuring the entanglement between the parties.
Finally, in Sec. \ref{heis}, we consider a specific form for the Hamiltonian $H$, 
that of a one-dimensional Heisenberg spin chain, and check explicitly, 
in this simple case, the validity 
of the \emph{locality conjecture} formulated in Sec. \ref{lines:local}. 
Some conclusions and possible developments of this approach are finally discussed in Sec. 
\ref{last}.

\section{INFORMATION AND MEASUREMENT: WORKING HYPOTHESES} \label{relational}

The description of the measurement process as a unitary evolution
in the direct product space of the ``measured" and the ``measuring" systems 
was first proposed by Everett in the celebrated work \cite{everett} which gave 
rise to the ``many-worlds interpretation'' of quantum mechanics.
Rather than to a ``many-worlds reality'' though, Everett's approach 
emphasizes the role of relative information in quantum mechanics 
and points to a ``relational reality'' \cite{rovelli}, 
where real and definite is in fact only the relative information that physical systems have about each other. Other useful insights along these lines are found in \cite{zurek,deu,mermin,poulin}.
We call this interpretative scheme of the quantum measurement "relational interpretation" (RI) and summarize it in Sec. \ref{ri}. For reasons that will be clear in the following, we will assume in the rest of the paper a modified version of RI, namely, we will only account for the information coming from quantum entanglement. 
and thereby will not consider classical correlations as a source of information. 
This is better explained in Sec. \ref{qrv}.

\subsection{The Relational Interpretation} \label{ri}

The relational interpretation (RI) is briefly reviewed here by taking as an example the measurement of an electron's spin. We follow Ref. \cite{rovelli}, that we recommend to readers looking for a deeper and more 
exhaustive analysis. Here we are not going to discuss 
decoherence \cite{zurek,deu}, that still plays a very important role in this framework.

Consider an electron in a generic spin state $ |\psi\ket\ \equiv\ v_+ |+\ket + v_- |-\ket $ with $|v_+|^2+|v_-|^2 = 1$. $|+\ket$ and $|-\ket$ are the two eigenvectors of the spin operator along the $z$ axis, along which the spin measurement is made.
The ``reduction of the wave packet" postulate says that, while measured,
the state $|\psi\ket$ gets projected  in either of the two eigenvectors,
$|+\ket$ and $|-\ket$, with probability $|v_+|^2$ and $|v_-|^2$, according to whether the outcome of the measurement has been ``spin up" or ``spin down" respectively. 
Therefore if we know that the measurement has been performed but we haven't read the outcome yet,  we describe the electron by means of a density matrix:
\begin{equation} \label{rel:dense}
\rho_{e}\ = \  |v_+|^2 \, |+\ket\bra+|\ +\ |v_-|^2 \, |-\ket\bra-| \ .
\end{equation}

In order to see how RI describes the same process we move from the Hilbert space
of the electron ${\cal H}_e$ to the larger Hilbert 
space ${\cal H} = {\cal H}_e \otimes {\cal H}_M$ containing also the degrees of freedom of the measuring apparatus $M$. According to RI
the apparatus $M$ is a quantum system as much as the electron and we assume for the moment that 
${\cal H}_M$ is a two-dimensional Hilbert space, 
spanned by the the two basis vectors $|+\ket_M$ and $|-\ket_M$. Such states correspond to the two possible outcomes of the experiment and are therefore associated to, say, "apparatus with the hand pointing towards spin up" and "apparatus with the hand pointing towards spin down" respectively. 
The measurement is now described by a unitary evolution in the tensor 
product space ${\cal H} = {\cal H}_e \otimes {\cal H}_M$ (say, from $t=-\infty$ to $t = +\infty$) that \emph{correlates} the state of the apparatus with that of the electron:
\begin{equation} \label{rel:evolution}
|\Psi (-\infty)\ket \ =\  |\psi\ket\  \otimes \ |{\rm init.}\ket_M \ \
\longrightarrow \ \ |\Psi (+\infty)\ket \ =\ v_+ |+\ket \otimes |+\ket_M \ + \ v_- |-\ket \otimes |-\ket_M 
\end{equation}

Why is the unitary evolution \eqref{rel:evolution} claimed to be a measurement?
We may note that the final state $|\Psi(+\infty)\ket$ traced over $M$'s degrees of freedom gives the same density matrix \eqref{rel:dense} for the system electron ``alone". But this is not the end of the story. 
In information theory the correlations between two systems are a measure of their mutual information. In reference \cite{rovelli} it is argued that the vectors belonging to the subspace spanned by  $|+\ket \otimes |+\ket_M$ and $|-\ket \otimes |-\ket_M$ are the information theoretical and quantum mechanical 
descriptions of the state of affairs "$M$ \emph{knows} the spin of $e$ along $z$" (and \emph{vice versa}: "$e$ \emph{knows} the position of the hand of $M$").
More formally, a linear operator ${\cal K}$ measuring the relative "knowledge" between $M$ and $e$ can be defined. With respect to the basis $\{ |+\ket \otimes |+\ket_M, \  |+\ket \otimes |-\ket_M,\   |-\ket \otimes |+\ket_M, \  |-\ket \otimes |-\ket_M\}$ it reads
\begin{equation} \label{rel:knowledge}
{\cal K}
 \ =\ 
\begin{pmatrix}
1&0&0&0\\
0&0&0&0\\
0&0&0&0\\
0&0&0&1 
\end{pmatrix} .
\end{equation}
The generally time-dependent expectation value 
\begin{equation} \label{rel:P}
P(t) \ =\ \bra\Psi (t) |\, {\cal K} \, |\Psi(t)\ket   \, .
\end{equation}
gives the probability that $M$ knows the spin of $e$ when the whole system 
is in the state $|\Psi(t)\ket$. 
By knowing the details of the evolution 
\eqref{rel:evolution} we may therefore follow the measurement  and at 
some intermediate step $t$  
ask ``does $M$ \emph{already} know about $e$?'' In this framework this is a 
genuine quantum (yes/no) question and therefore it has a probabilistic (yes/no) answer: the probability that $M$ knows already about $e$ is \eqref{rel:P} \cite{rovelli2}.
At the end of process \eqref{rel:evolution} $P(+\infty) = 1$ and therefore we are sure that the apparatus eventually \emph{knows} the value of the spin of the electron. 
 Of course, we can not tell which of the two possible outcomes ($+$ or $-$) has been actually measured by $M$ although, by knowing the values of $v_+$ and $v_-$, we can tell the \emph{probability} with which $M$ has measured, say, ``$+$''. Note that there is no point in asking "what's the spin of the electron after the measurement" because according to RI, there is no collapse of the wavefunction and therefore the spin of the electron has not, in general, any definite value. What is definite -- but not accessible to anybody external to the electron-apparatus system -- is only the information that the apparatus has about the spin of the electron. 

We can now take a  look at the hand of the apparatus and see whether it has 
measured ``$+$'' or ``$-$''. In order to do this we need to \emph{interact} with the 
apparatus and ``measure'' it. Again, this process can be described 
``from the outside'' as a unitary evolution. The final state of the experimenter (say, ``ME'') who reads the outcome will be
correlated with the apparatus and therefore also with the electron. Schematically, in the tensor product space $\cH_e\otimes\cH_M\otimes\cH_{ME}$, we will have
\begin{equation} \label{rel:evolution2}
v_+ |+\ket \otimes |+\ket_M \otimes |\text{ME reading ``$+$''}\ket \ + \ v_- |-\ket \otimes |-\ket_M \otimes |\text{ME reading ``$-$''}\ket
\end{equation}
This is an information theoretical description of the state of affairs "I know what the apparatus has measured" and also "I know the spin of the electron".
Of course, because of the correlation that has been created, our knowledge about 
the spin of the electron (either ``$+$'' or ``$-$'') 
is consistent with the answers that we would get by measuring it again and again 
with, say, other measuring apparatuses; exactly \emph{as if} the 
electron had collapsed in either of the two states $|+\ket$ or $|-\ket$. 

\subsection{Entropy and quantum information}\label{qrv}

What is borrowed here from the relational view (RI) is summarized in the
following body of assumptions: 

\begin{quote}
(I) \emph{Relational Assumptions:} Every physical process is described by a unitary evolution of a state vector in a Hilbert space.  A measurement is a physical process  
which increases the degree of correlation/information between the 
subsystems of ``the measuring" and ``the measured".  
The actual informational \emph{content} of a physical process 
is intrinsically relational i.e. relative only to 
the subsystems taking part in the process. Physical theories can only predict the probability associated to a given informational content. 
\end{quote}

By taking the relational view (RI) one may be tempted to re-interpret a given dynamical evolution as described e.g. in some realistic quantum field theory as measurements actually going on between different parties. In more general frameworks, however,
defining a knowledge operator  such as ${\cal K}$  \eqref{rel:knowledge} is very subtle. In the last subsection we assigned the state vector $|+\ket_M$ to that particular condition of the measuring apparatus that we called "apparatus measuring $+$". That was a very relevant interpretative step: we were in fact relating a huge number of microscopic degrees of freedom with the macroscopic condition "the hand of the apparatus points toward $+$". Imagine that we want to deal only with microscopic degrees of freedom, e.g the spin of an electron ($e$) and the isospin of a nucleon ($n$). Then it looks very hard to tell why the state $|+\ket_e \otimes |+\ket_n$ should describe relational ``knowledge" and, on the opposite, $|+\ket_e \otimes |-\ket_n$ should describe complete relational "ignorance". In other words, a knowledge operator such as \eqref{rel:knowledge} is always defined with respect to the bases chosen in the Hilbert spaces of the two parties; but, when dealing with \emph{bona fide}-fundamental degrees of freedom, we just don't know what meaning to give to any particular basis chosen. 

A basis-independent type of correlation is 
\emph{quantum entanglement}.
Finding a general measure of entanglement is still an open problem in quantum information theory \cite{vedral,adami}.  
If a system $A$ is described by a density matrix $\rho_A$, 
the \emph{von Neumann entropy} $S$ is a definite positive quantity defined as
\begin{equation} \label{lines:von}
S(A) \ =\ - {\rm Tr}_A (\rho_A \log_2 \rho_A). 
\end{equation}
If $A$ belongs to a bipartite system $AB$ that is in a pure state, i.e. $\rho_{AB} = |\Psi\ket\bra\Psi|$, $\rho_A = {\rm Tr}_B |\Psi\ket\bra\Psi|$, then the von Neumann entropy $S(A)$ is a very good measure of the entanglement between $A$ and $B$. In this case, moreover, $S(A) = S(B)$. If there is no quantum correlation between the two subsystems then $S(A)=S(B)=0$. This is the case of a \emph{factorized} or \emph{separable} state i.e. a direct product state such as
$ |\alpha\ket_A \otimes |\beta\ket_B$. 
Maximally entangled states, on the opposite, have maximal entropy. Take, e.g., the state vector
\begin{equation}
\frac{1}{\sqrt{D}}\sum_{j=1}^D |a_j\ket_A \otimes |b_j\ket_B \, ,
\end{equation}
where $D$ is the dimension of both ${\cal H}_A$ and ${\cal H}_B$ and $ |a_j\ket_A$ and $|b_j\ket_B$ are two ortonormal bases. In this case we have $S(A) = S(B) = \log_2 D$. This is also the 
maximal amount of information (number of bits) that a $D$-dimensional quantum system can have about the ``outside". However, in most situations, the system $AB$ is itself a part of a larger system $U$ and therefore not describable, in general, as a pure state. In this case, the correct mathematical expression
for the mutual entanglement of $A$ and $B$ is still a matter of debate.
Among many proposals \cite{various} we do not choose any 
in particular and just call $I(A,B)$ our \emph{bona fide} measure of 
mutual entanglement between $A$ and $B$; being symmetric under the 
exchange between $A$ and $B$ and reducing to the von Neumann entropy 
$S(A)=S(B)$ when $AB$ is in a pure state are both basic necessary 
conditions for a good definition of $I(A,B)$. When possible,  
we stay on the ``safe side'' and deal only with bipartite systems 
in a pure state and with the von Neumann entropy $S$.
For instance, in the calculations of Sec. \ref{model} we 
consider only the entanglement between some given subsystem $A$ and the rest of the Universe, which is assumed, as a whole, to be in a pure state; in this case the von Neumann entropy $S(A)$ (or the \emph{linear entropy} to be introduce thereafter) is certainly a good measure of entanglement.

For  definiteness, we will assume that the mutual entanglement $I(A,B)$ is an extremely good indicator of the actual amount of information shared between $A$ and $B$. In other words, the probability $P(n)$ that $A$ and $B$ share $n$ bits of information is a distribution very
sharply peaked around the value $I(A,B)$.
This and the other above observations, which constitute the main shortcut with respect to the RI scheme, are summarized in the following. 
\begin{quote}
(II) \emph{Entanglement Assumption:} At a microscopic/fundamental level entanglement is the only source of mutual information. The mutual entanglement $I(A,B)$ measures the amount of relative information between $A$ and $B$. Such a relative information can be encoded in a string of $n$ bits, $n$ being either of the two integers closest to the real number $I(A,B)$.
\end{quote}

\section{WORLD-LINES OF INFORMATION AND LOCALITY}

\label{lines}

The importance of the information theoretical content of physical theories has been most authoritatively stressed by John Wheeler  in his ``it from bit" program: ``every \emph{it} -- every particle, every field or force, even the spacetime continuum itself derives its function, its meaning, its very existence entirely from the apparatus-elicited answers to yes-or-no questions, binary choices, \emph{bits}" \cite{wheeler}. 
In the last section we argued that, within the usual formalism of quantum mechanics,
such "answers to yes-no questions" are what physical subsystems actually give to each other during the deterministic and unitary evolution of their state vector. In this section we try to infer under which circumstances and in which sense a sort of spacetime description can be given to such an abstract and pre-geometric information exchange.
Occasionally, we  take the point of view of a physical system (``\emph{Vincent}" or just ``$V$") that, because of its interactions with other systems, receives information from and about ``the outside". 

\subsection{Vincent and the others} \label{vincent}

\emph{Vincent} is a subsystem and therefore is represented by a factor in the choosen tensor product structure (TPS) of the Hilbert space of the Universe, say 
\begin{equation} \label{lines:friends}
\cH_{\rm Universe} = \cH_{V} \otimes \cH_A \otimes \cH_B \otimes \cH_C \otimes \dots\ . 
\end{equation}
The ``outside'' are the other factors. We assume that \emph{Vincent} is big and composite enough to store every bit of information received. Of course, one may naively think of \emph{Vincent} as a human being on a spacetime journey that writes all its information on a piece of paper, although, according to the assumptions of  Sec. \ref{relational}, every physical system receives and stores information. 

The central issue of this section is to characterize, among all possible TPSs,  the ones that single out objects/subsystems "localized in space". We start by formulating a reasonable and generic  \emph{necessary} condition:
\begin{quote}
(III) \emph{Finite Information Assumption:} Any ``localized" and ``finite size" system can share with the rest of the Universe up to a finite amount of information.
\end{quote} 
This property (which presumes a sort of informational UV cut-off \cite{kempf}) will be discussed at the end of Sec. \ref{toy} and is assumed to apply to \emph{Vincent} and to the other parties $A, B,$ etc\dots\  
By working in the Schroedinger picture we can follow the evolution of the state vector of the Universe $|\Psi(t)\ket$ as predicted by some physical theory (i.e. some Hamiltonian operator $H$) and tell the amount of the information exchange between the subsystems \eqref{lines:friends}. According to assumption (II) 
the relevant quantities to be taken into account are the (time dependent) mutual entanglements 
\begin{equation} \label{lines:mutual}
I(\text{\emph{V}, rest of the Universe};\, t) \, =\, S(V;\, t), \quad I(\text{\emph{V, A}};\, t), \quad {\rm etc} \dots 
\end{equation}
obtained from the density matrices 
\begin{equation} 
\rho(t) = |\Psi(t)\ket \bra \Psi(t)|, \quad
\rho_{ VA}(t) = \text{Tr}_{\, \text{all but \emph{VA}}}\, |\Psi(t)\ket \bra \Psi(t)|, \quad {\rm etc}\dots 
\end{equation}
Of course, since the state vector of the Universe $|\Psi(t)\ket$ evolves in 
a unitary and reversible way, mutual entanglements can increase as well as decrease in time and 
information can also be lost. In low dimensional quantum systems, for instance,
entanglement typically undergoes comparatively large oscillations. Here we
assume that all the mutual entanglements $I(V,A;t),\, I(V,B;t),$ etc\dots\  
are non-decreasing functions of time\footnote{Once created, entanglement is indeed expected to be a very general feature of the future evolution,
since the volume of non-entangled states is exponentially small in the 
dimensions of the total Hilbert space $\cH_{\rm Universe}$ \cite{zyc}.}; the admittedly prejudicial picture we have in mind is that 
of a sufficiently large Universe where diluted subsystems ``meet'' each other once and never meet again. Think at the scattering between two particles whose states are initially separable and get entangled during the scattering process. In principle, the particles may as well disentangle during their further evolution in a sort of
"rewind scattering" but, in practice, this is very unlikely to happen. 
Again, in a way, this assumption may be thought as a necessary requirement for the TPS \eqref{lines:friends} in order to interpret the subsystems $V$, $A$, $B$, etc\dots\ 
as ``localized objects''.

Say that, between two instants $t_1$ and $t_2$, these are the bits of information received by \emph{Vincent}: 
\begin{equation} \label{lines:string1}
110001010011101 01000110 000010011100011 01000110 101\, ,
\end{equation}
and that before $t_1$ \emph{Vincent} is in a pure state \emph{i.e.} $S(V;t<t_1) = 0$.
Few comments are in order:

($i$) According to assumption (I), the string of bits \eqref{lines:string1} is generally not predictable with certainty by any physical theory. All we can predict by solving the dynamical evolution 
for the state vector of the Universe $|\Psi(t)\ket$ 
is only the \emph{probability} that \emph{Vincent} actually gets some given string of bits. On the other hand, according to assumption (II), the \emph{number} of bits \eqref{lines:string1} is given, to a very good approximation, by the quantity 
$S(V, t_2) - S(V,t_1)$. 

($ii$) The time parameter $t$ controls the evolution of the \emph{probabilities} that we associate to something that actually "happens" to \emph{Vincent}. As with the (linear) knowledge operator \eqref{rel:knowledge} in the RI
scheme, we can use here the (non linear) von Neumann entropy $S(V;t)$ to tell the probability that, at some instant $t$, \emph{Vincent} has already received, say, his $n^{th}$ bit of information from the outside\footnote{Unfortunately, in the litterature on quantum information theory, we have not found about the amount of quantum information--probability distribution. 
We are assuming here that such a probability distribution exists and that it is very sharply peaked around its average value $I(A,B)$ (see assumption II). 
Therefore, the question ``has \emph{Vincent} already received his $n^{th}$ bit of information'' can be given a precise probabilistic answer by the theory.}.

($iii$) At some given instant $t$, \emph{Vincent} has \emph{either} already 
received some bit of information \emph{or} not received it yet. In this respect,  
from \emph{Vincent}'s perspective, physical processes are genuinely ``quantized''.
At any step, \emph{Vincent} can, ideally, look back at the information that
it has received so far and therefore reconstruct the string of 
bits \eqref{lines:string1} in its correct order.

\subsection{Coincidence} \label{cont}

The elementary observation at the basis of this paper is that
every assertion about a spacetime is in fact an assertion about the mutual geometric relations between physical systems ``living" in such a spacetime.
The most elementary of such relations is \emph{coincidence}, what, in a spacetime language, can be translated as ``being in the same place at the same time". Other relations, such as ``being at a certain distance" or ``being separated by a certain time interval", are more involved because they presume the concept of coincidence in their operational definitions. One can measure, for instance, the distance between two objects only after these objects have been made coincident with the ends of a ruler or, in interferometry experiments, with the reflection of an electromagnetic wave. 

From a pre-geometric perspective, it looks natural to define coincidence by means of physical interactions, i.e. to define  two parties as ``having been coincident" if they ``have physically interacted" with each other. The idea is pretty intuitive and generally comforted by
the known local character of physical laws\footnote{By looking at it the other way around one may speculate that
interactions (terms in a Lagrangian) are ``local" ultimately because physical interactions (physical processes) are the only way through which we perceive and operationally define ``coincidence" and ``locality".}, although it may seem hard to make it   
any less vague within the usual formalism of quantum mechanics:
an elastic scattering between two particles, for instance, is a good intuitive example of a ``localized physical interaction" but, when described in terms of wave functions, it may well be spread over all spacetime.
Furthermore, by considering higher and higher values of the impact parameter (say, up to 1 light-year), the same formalism ends up describing as a ``scattering" something which is not localized at all and that can even hardly be considered as a ``definite physical process". 
On the opposite, in any experimenter's experience, if a photographic plate gets imprinted by a particle, this always happens in a very precise point at a very precise moment; if the Geiger counter clicks, this is a very definite and localized physical process. 

We have argued that the common experience of physical processes as ``something definite and happening" and the theoretical continuous Hilbert space description of them are not in contraddiction when considered on the basis of information. The string of bits \eqref{lines:string1} is how the system/observer \emph{Vincent} sees the world; at any time its (ideal) Geiger counter has either already clicked or not clicked yet for the simple reason that the information about the outside cannot arrive in amounts smaller than one single bit. In this scheme, therefore, the coincidence between two systems
can be given the precise meaning of ``information exchange", the 
mutual entanglement being the measure of the probability  that such information 
exchange has indeed happened. 
As regards the above example of two scattering particles we can argue that, for huge 
values of their impact parameter, the mutual entanglement
will never exceed some very thin fraction of unity and that 
the probability that either system has been coincident with the other is therefore extremely low.
Again, what we are going to define as ``coincidence'' is an 
intrinsically relational concept \cite{rovelli}.

Before attempting a more precise operational definition of coincidence it is 
worth considering a counterexample in which the increase of entanglement between 
two systems does not reflect at all the intuitive idea of ``having been in the same 
place at the same time''. 
Let's go back to \emph{Vincent} and its neighbours and suppose that, before 
some instant $t_1$, the system $A$ is already entangled
with system $B$. For definiteness, one may think of $A$ and $B$ as 
two two-dimensional systems very far from each other and in the typical EPR state 
$(|+\ket_A\otimes|-\ket_B - |-\ket_A\otimes|+\ket_B)/\sqrt{2}$. 
Since $A$ and $B$ are maximally entangled \emph{Vincent} cannot
get entangled with $A$ without, at the same time, getting entangled 
also with $B$\footnote{Our interpretative scheme looks
consistent with the usual EPR argument, according to the which, after measuring $A$, 
we are able to predict the outcome of a measurement made on $B$ i.e. we also get 
information about $B$:
measuring $A$ implies getting entangled with $A$ and, \emph{a fortiori}, with $B$.}.
Therefore, if $I(V,B;t)$ grows in time since $t_1$ of an amount of order unity this 
does not necessarily mean that \emph{Vincent} has been coincident with $B$; 
on the opposite, \emph{Vincent} may have been ``in touch'' with $A$ instead! 

In order to isolate the parties that have effectively been ``in touch'' one should check that no information has been exchanged with any other party. In order to accomplish this, one may attempt to define  \emph{Vincent} as ``having been coincident with $A$ between $t_1$ and $t_2$'' in the case that $I(V,A;t_2) -  I(V,A;t_1) =    S(V;t_2) -  S(V;t_1) = S(A;t_2) -  S(A;t_1)$. However, unfortunately, we cannot count on a general and quantitatively reliable expression of the mutual entanglement $I$; in order to stay on the safe side, we restrict our definition
of coincidence to systems which are initially in a pure state,
say, $|\Psi\ket_{\rm Universe} = |\ket_A \otimes |\ket_B \otimes 
|\ket_{\text{everything else}}$ and just formulate a sufficient condition: 
\begin{quote}
(IV) \emph{Spacetime coincidence (sufficient condition)}: If before the instant $t_1$ the systems $A$ and $B$ 
are in a pure state \emph{i.e.}  $S(A;t<t_1) = S(B;t<t_1) =0$ and, at a later time $t_2$ they are entangled, 
$S(A;t_2) = S(B;t_2) > 0$, without, during all the process, having mixed with anything else, $S(AB;t<t_2) = 0$,
then $A$ and $B$ \emph{have been coincident} with each other between $t_1$ and $t_2$.
\end{quote}
The above condition is very strict and probably improvable. As it stands, \emph{Vincent} can 
define its coincidence only with the very first system it gets entangled with since, afterwards, \emph{Vincent} is not in a pure state any longer.
Nevertheless, according to assuption IV,
by looking back at its ``informational wordline'', 
\emph{Vincent} can legitimately interpret, say,
\begin{equation} \label{lines:string2}
\underbrace{1100010100}_{\text{``I met $A$''}}11101 01000110 000010011100011 01000110 101\, ,
\end{equation}
Whether or not \emph{Vincent} is in fact clever enough to interpret those bits 
and recognize them as 
coming from system $A$ is not important here. In \eqref{lines:string2} we have 
just drawn the  interpretation that a sufficiently clever system may legitimately 
give of its stored information. 

As a last remark we note that if two systems $A$ and $B$ have been coincident between $t_1$ and $t_2$ then coincidence generally applies also to many other ``larger" systems containing $A$ and $B$ as subsystems. 
Say, for instance, that two other systems $D$ and $E$ are pure and do not mix with anything else during the same time interval \emph{i.e.} $S(D;t<t_2) = S(E;t<t_2) =0$; if coincidence definition 
(IV) applies to the couple $A$ and $B$ it trivially applies also to the couple of larger systems $AD$ and $BE$. The opposite can also be true. Say that $A$ is itself a composite systems \emph{i.e.} $\cH_A = \cH_{A1}\otimes\cH_{A2} \otimes \dots$; we may discover that the subsystem $A2$ is in fact ``responsible" at a deeper level for the coincidence between $A$ and $B$ or, in other words, that coincidence applies to the couple $A$ and $B$ but applies also to $A2$ and $B$. The ``smaller" are the systems to which coincidence is recognized to apply the finer grained is the spacetime description that we are able to give.

\subsection{Vincent's time} \label{time}

The presence of a ``global" time parameter $t$ in our formulas may 
look like a step-backward when compared to the usual relativistic-invariant formulation 
of physical theories. 
The hope, however, is to recover \emph{a posteriori} the complete general covariance 
just as in the case of the Hamiltonian formulation of field theories, 
where a split of spacetime into ``space'' and ``time'' is always required at the beginning. Worse than this, the introduction \emph{ab initio} of a time parameter may
seem at odds with our pre-geometric pretensions.  Admittedly, in this scheme, we need from the beginning an external time parameter $t$. The latter, however, has merely the function of ordering (on probabilistic grounds) the fluxes of information between the subsystems and has in principle nothing to do with the time as 
perceived and defined by the observers. In order to clarify this issue 
we provide \emph{Vincent} with a clock $C$, another physical system that regularly sends pulses of information to \emph{Vincent}. By looking at its information string \eqref{lines:string1}, \emph{Vincent} can try to recognize the pulses of the clock  and give a temporal description of its recent history. A good clock must be, therefore, at least
recognizable; 
each pulse should have an extremely high probability to 
be sent entirely within a sharp time interval $\delta t$ 
and to carry a precisely encoded informational content (say, for simplicity, always  "$01000110$");
looking at the string \eqref{lines:string1} \emph{Vincent} may thus (conventionally) 
interpret:
\begin{equation} \label{lines:string3}
\underbrace{1100010100}_{\text{``I met $A$''}} 11101 \underbrace{01000110} _{\text{``it was 8"}} 000010011100011 \underbrace{01000110} _{\text{``it was 9"}} 101\, ,
\end{equation}
A good clock should also be able to give a good account for all the phenomena 
that are \emph{bona fide} periodic for \emph{Vincent}. In other words, the arrival
of the pulses should, in turn, be itself predictable with some precision
on the basis of \emph{Vincent}'s experience. Predictable, in this scheme, means being 
in some precise relation with the external time $t$ although such a relation 
may be whatever; for instance, there is no need
for each pulse to happen at every fixed time interval $\Delta t$. The external time $t$ is 
therefore not 
accessible to \emph{Vincent} nor to any other subsystem in the Universe and has merely the role of ordering (on probabilistic grounds) the information fluxes between the parties. 
Slightly different ideas along the same lines are found in the 
works \cite{page1,page2,banks}, where
the ``internal'' time for the system--observer is defined through a relational scheme 
which has been partially emulated here and where the ``inaccessible'' character
of the external ``Schroedinger''  time $t$ has been first pointed out. 

Assertions such as ``I met $A$ before 8 o'clock", 
are already a primitive and elementary type of spacetime description, 
where the "informational worldline" of a system has been interpreted 
at the basic level of its ``past contiguities'' with other parties. Again, such 
a description may be enhanced as soon as a more general definition of ``coincidence'' 
is found that applies also to systems which are already entangled with something else.

\subsection{Localized systems} \label{lines:local}

In quantum mechanics tensor products are most often used to \emph{construct} the Hilbert space of a composite system once the spaces of the components are known. However, one may also go the other direction
and see in how many ways a given Hilbert space can be \emph{decomposed} into ``virtual subsystems" \cite{paolo1} or, in other words, in how many ways a Hilbert space can be given a tensor product structure (TPS). From a pre-geometric perspective, what to consider as a ``subsystem" is purely a matter of convention and the freedom in choosing a TPS embarassingly huge. 
For instance, one may choose as a physical system the pair $A)$ an electron on Mars and $B)$ a neutrino on the Sun; but one may as well weirdly mix these degrees of freedom and deliberately decompose the same system into $A')$ spin of the electron + elicity of the neutrino and $B')$ position of the electron + position of the neutrino, ending up with a completely non-localized description. Once the dynamics of the system-Universe is assigned, moreover, there is no particular reason for not considering also TPSs
depending on the ``external" time parameter $t$.

In order to quantify our \emph{a priori} freedom of choosing a TPS consider the example of a Hilbert space $\cH$ of dimensions $D=d^N$. This can always be decomposed in a tensor product $\cH_d^{\otimes N}$ of $N$ Hilbert spaces $\cH_d$ each of dimension $d$. Call $\{ 
 |1\ket_d, |2\ket_d, \dots, |d\ket_d \}$ a basis of $\cH_d$. In order to pick up
one of such possible decompositions we need to choose a way to identify
a given basis of $\cH$, say, $\{ |1\ket, |2\ket, \dots, |D\ket \}$, with the natural basis of $\cH_d^{\otimes N}$, 
\begin{equation}
\{ \underbrace{|1\ket_d \otimes|1\ket_d \otimes \dots \otimes |1\ket_d}_{\text{$N$ times}}, \ \ |1\ket_d \otimes|1\ket_d \otimes \dots \otimes |2\ket_d,\ \  \dots\ , \ |d\ket_d \otimes|d\ket_d \otimes \dots \otimes |d\ket_d
\}\, .
\end{equation}
Note also that, once a particular correspondence between these two bases have been choosen, a further change of basis inside each component $\cH_d$ does not change the TPS.
Thus, choosing a particular TPS of the type $\cH_d^{\otimes N}$ on $\cH$ amounts to choosing an element of the group $U(d^N)/U(d)^{ N}$ whose dimensions grow as $\sim d^{2 N}$.

The notion of coincidence between systems introduced in Sec. \ref{cont} has not much to do with the idea of locality since it can be applied to completely non-localized subsystems choosen from an arbitrary (and possibly time $t$ dependent!) TPS. Moreover, we have argued that, once coincidence applies to systems $A$ and $B$, it will most probably apply also to a pletora of pairs of larger -- and less localized -- systems containing $A$ and $B$ as subsystems. 
It is pretty intuitive, however, that the less ``localized" are the subsystems 
the more probably they will get in touch with each other during the evolution. 
By using such a geometric intuition and by trusting as physically sensible
the operational definition (IV) of coincidence we are lead to
formulate the following generic conjecture:
\begin{quote}
(V) \emph{Locality Conjecture:} 
\emph{Localized systems} have the less tendency to create \emph{coincidence} relations with each other: the (possibly time-dependent) tensor product structure that singles out \emph{localized systems} is the one in which the entanglement of initially completely factorized states \emph{minimally} grows during time evolution. 
\end{quote}

The above conjecture faces the risk of being made meaningless
by the huge freedom of TPS choices. In other words, there may always be a
suitable time dependent TPS that totally 
reabsorbs the effects of any given dynamics in such a way that the ones that 
we define as ``localized'' are in fact systems 
that never get entangled and do not exchange information at all. 
In this case there would be no coincidence relations between any localized system and 
therefore no spacetime relations at all. 
In the next section we argue that the structure of second quantized quantum systems
dramatically constrains the freedom of TPS choices. 
For free theories all 
the dynamics can in fact be reabsorbed by a suitable TPS choice and, therefore, spacetime 
cannot be given any operational meaning. On the contrary, in the case of 
interacting theories, the \emph{locality conjecture} can be applied in a non-trivial way.

\section{THE IDEA AT WORK} \label{model}

In order to apply the above ideas we consider in this section a general 
second-quantized interacting model with $N$ fermionic degrees of freedom
and analyze its informational content. 
The basics of the model and the Fock structure of its total Hilbert space $\cH$ induced by the
$N$ fermionic creators $\cd_j$ are briefly introduced in Sec. \ref{toy}. Since $\cH$ is
$2^N$--dimensional, the most fine grained TPS structure is the one made by $N$ 
two-dimensional parties. 
In Sec. \ref{parties} it is argued that the Fock structure dramatically reduces
the freedom of choosing a TPS on $\cH$ and a class physically
sensible TPSs is discussed. The first of them is the ``\emph{defining}--TPS'': the one
related to the occupation number representation and directly induced by the fermionic 
creators $\cd_j$; the others are obtained from the defining--TPS by applying
(possibly time-dependent)
Bogoliubov transformations. In Sec. \ref{def} the general instruments for calculating general
bipartite entanglements in this model are given; thereafter, we restrict to calculating 
the entanglement between one of the $N$ parties and the rest of the system--Universe. 
In Sec. \ref{model:time} 
the time evolution of the vector states in three cases of interest is analysed.
Sec. \ref{model:evolution} deals with the \emph{entanglement's evolution} with respect to
time-dependent TPSs of time-dependent state vectors. In Sec. \ref{heis} we check
conjecture (V) in 
the case of a one-dimensional Heisenberg spin chain. We prove that the TPS traditionally 
associated with position in fact minimizes the growth of entanglement.

\subsection{The toy model} \label{toy}

The Hamiltonian of this toy model describes $N$ fermionic degrees of freedom; it has a free and 
a generic four fermion interaction part,
\begin{equation} \label{model:hamiltonian}
H \ =\  H_0 + H_I \ =\ \sum_{j } \lambda_j \, \cd_j c_j\  + \ \sum_{jklm} \g_{jklm} \, \cd_j \cd_k c_l \, c_m\, ,
\end{equation} 
where $c_j$ are the usual fermionic annihilators, satisfying the anti-commutation relations
\begin{equation}\label{model:anti}
\{\cd_j, c_k\}\, = \, \delta_{j k}, \qquad  
\{c_j, c_k\}\, = \, 0 \, .
\end{equation}
In virtue of the above relations, without changing the Hamiltonian it is always possible to choose a ``maximally symmetric" kernel $\g$:
\begin{equation} \label{model:sym1}
\g_{jklm}\ =\ -\ \g_{kjlm}, \qquad \g_{jklm}\ =\ -\ \g_{jkml}\, .
\end{equation}
Moreover, since $H$ is self-adjoint, 
\begin{equation} \label{model:sym2}
\lambda_j = \lambda_j^*\, , \qquad\g_{jklm}^* = \g_{mlkj}\, .
\end{equation}
Note also that any free Hamiltonian $H_0$ quadratic in the fields can be brought into the diagonal  ``\emph{momentum basis}'' form of eq. \eqref{model:hamiltonian} by means of a Bogoliubov transformation. For definiteness we also assume that $\lambda_j \geq 0$ for every $j =1\dots N$. This can always be achieved by a ``switch re-definition" between the
creator and annihilator e.g.
\begin{equation}
H_0 \ =\  \cd_1 c_1 - \cd_2 c_2 \ =\  \cd_1 c_1 + \cb^\dagger_2 \cb_2 \qquad (\cb_2 \equiv \cd_2) 
\end{equation}
This guaranties that the state annihilated by all $c_j$ is the Hamiltonian eigenstate with minimal energy: the vacuum $|0\ket$.

The big advantage in dealing with fermionic degrees of freedom is that the 
Hilbert space ${\cal H}$ of our toy-Universe is finite dimensional, 
${\rm dim}({\cal H}) = 2^N$. By starting from the vacuum $|0\ket$ and repeatedly 
applying the creator operators one recovers
the usual Fock structure
\begin{equation} \label{model:fock}
{\cal H} = \cH_0 \oplus \cH_1 \oplus \cH_2 \oplus \dots \oplus \cH_N \, , 
\end{equation}
$\cH_M$ being the subspace ``with $M$ particles", of dimension $N$-choose-$M$:
\begin{equation}
2^N\ =\ \text{dim}( \cH)\ =\ \sum_{M=0}^N \text{dim} (\cH_M)\ =\ \sum_{M=0}^N \binom{N}{M} .
\end{equation}
The Fock structure is particularly relevant in this simple second quantized model since the number operator $n = \sum_i \cd_i c_i$ commutes with the Hamiltonian and the particle number is strictly conserved during the evolution. 
A generic vector belonging to the $M$-particles subspace $\cH_M$ is associated to a totally anti-symmetric tensor $v^\M_{p_1\dots p_M}$ with indeces $p_j$ running from $1$ to $N$,
\begin{equation} \label{model:mpar}
|\psi,\ \text{$M$ particles}\ket\ =\ \sum_{p_1 \dots p_M} v^\M_{p_1\dots p_M}\ \cd_{p_1} \dots \cd_{p_M} |0\ket\, ,
\end{equation}
where $v^\M_{p_1 p_2 p_3 \dots p_M} = - v^\M_{p_2 p_1 p_3 \dots p_M} = v^\M_{p_2 p_3  \dots p_M p_1}$ etc\dots and, in order for the state vector to be properly normalized,
\begin{equation} \label{model:normv}
 \sum_{p_1 \dots p_M} |v^\M_{p_1\dots p_M}|^2\ =\ \frac{1}{M!}\, .
\end{equation}
Any generic vector of $\cH$ can be expressed as a normalized combination of 
the state vectors \eqref{model:mpar} of fixed numbers of particles. However, 
because of the number of particles conservation, the different subspaces 
$\cH_0, \cH_1, \dots, \cH_N$ never ``talk to each other''. Since this system 
pretends to describe an entire ``Universe'', 
or at least to be isolated from everything else,
we are in the presence of a 
superselection rule and we can therefore write the generic state 
through a  block diagonal density matrix:
\begin{equation} \label{model:block}
\rho \ =\ \sum_{M=0}^N \, p_M\, |\psi,\ \text{$M$ particles}\ket 
\bra \psi,\ \text{$M$ particles}|, \qquad \text{where}\qquad  \sum_{M=0}^N p_M =1\, .
\end{equation}

Before going on and study the informational content of this toy model let's briefly discuss its main limitations when it is called to represent more realistic quantum theories, namely, the conservation of the particles number and the finite size of the total Hilbert space. 

The first of these features greately enhances the Fock structure \eqref{model:fock} and
allows to analyse separately the dynamics inside each fixed number of particles subspace.
Moreover, as argued in Sec. \ref{parties}, this feature
remarkably restricts the possible choices of a TPS. Admittedly, when trying
to generalize these ideas to more realistic and/or interesting
theories one should look for other conserved quantities and 
superselection rules. 

As regards the second point -- the finite dimensionality of our total Hilbert space -- 
one may argue 
that the sophistications of the Hilbert spaces that quantum field theories usually
deal with may originate from the ``prejudice'' of a pre-existing spacetime 
continuum. In the dreamland of pre-geometric physics, associating a quantum oscillator to 
each one of such a 
non-countable infinity of points may ultimately prove to be an overcomplication. 
A finite dimensional Hilbert space is, after all, what originates from a fermionic 
theory with an arbitrarily high UV cut-off and formulated in a finite size universe 
(\emph{i.e.} with a IR cut-off as well). 
In bosonic theories, on the other hand, it is true that an infinite dimensional Hilbert space 
is associated to each mode; 
but, again, in the presence of a UV cut-off, those energy levels higher than 
the cut-off can never be excited and therefore the effective total Hilbert space can be 
also argued to be finite dimensional.

\subsection{Different parties}\label{parties}

A very useful representation for the state vectors is the \emph{occupation number representation} 
\cite{bjork}. In other words, we can take as a basis for $\cH$, the vectors
\begin{equation} \label{model:productbasis}
|\alpha_1\ket \otimes |\alpha_2\ket \otimes \dots \otimes |\alpha_N\ket \ \equiv
\ \cd_N^{\, \alpha_N}\, \cd_{N-1}^{\, \alpha_{N-1}}\, \dots \cd_1^{\, \alpha_1}\, |0\ket,
\end{equation}
where the indexes $\alpha_p$ run from $0$ to $1$, and write a generic state vector of $\cH$ as 
\begin{equation} \label{model:psi}
|\psi\ket \ =\ \psi_{\alpha_1,\, \alpha_2,\, \dots \, ,\, \alpha_N}  \,
|\alpha_1\ket \otimes |\alpha_2\ket \otimes \dots \otimes |\alpha_N\ket , 
\end{equation}
where
\begin{equation} \label{model:normpsi}
\psi^*_{\alpha_1,\, \alpha_2,\, \dots \, ,\, \alpha_N}\, 
\psi_{\alpha_1,\, \alpha_2,\, \dots \, ,\, \alpha_N}\ = \ 1 .
\end{equation}
Note that a summation over repeated binary greek indices $\al$, $\beta$, etc\dots 
is always assumed in this section. On the contrary, latin indexes, 
$j, k, q, p$ etc\dots \,
run from $1$ to $N$, and the sum symbol over them is always explicitly written
when needed.

The occupation number representation \eqref{model:productbasis} directly induces a natural 
and finest grained TPS for the total Hilbert space $\cH$, made
of $N$ direct products of 
two dimensional Hilbert spaces $\mathbb{C}^2$ \cite{paolo2},
\begin{equation} \label{model:tps}
\cH \ =\ \underbrace{\mathbb{C}^2 \otimes \mathbb{C}^2 \otimes \, \dots\,  \otimes
\mathbb{C}^2}_{\text{$N$ times}}\, .
\end{equation} 
We call such a particular tensor product structure the ``\emph{defining}--TPS''. 
According to the remarks 
of Sec. \ref{lines:local} one can start from the defining--TPS and, by 
applying an element of the group $U(2^N)/U(2)^N$, move to any other possible 
$ {\mathbb{C}^2}^{\otimes N}$
TPS. Of course, to any of these new TPSs are still associated  
creators and annihilators operators $\ctd_j$ and $\ct_j$; in other words, 
the generic totally factorized vector can still be given in the form 
\eqref{model:productbasis} with the
$c$ operators replaced by some new $\ct$ operators. The new operators, however, will in 
general be defined with respect to the old ones in a very weird and possibly non-linear way and
the original Hamiltonian \eqref{model:hamiltonian} will definitely loose its direct 
interpretation of ``two fermions + four fermions interection''. Moreover, the new
``vacuum'' will be neither of minimal energy nor even stable under time evolution.
We argue that the second quantized formalism 
restricts the physically sensible TPSs to those 
that can be associated to new creators and annihilators that 
are in a linear relation with the old ones; or, in other words, that are connected 
to the old ones by means of a Bogoliubov transformation. Moreover, since in this case 
the number of particles is conserved, we require the Bogoliubov transformations to
strictly preserve the Fock structure \eqref{model:fock} and do not mix states with a 
different number of particles \emph{i.e.} do not mix creators with annihilators.
Our most general TPS is therefore induced by the new $\ctd_j$ and $\ct_j$ defined 
through
\begin{align}
\ctd_{q}& \  = \ \sum_p B^*_{p\, q}(t)\ \cd_p\, , \qquad \qquad \ct_q \  = 
\ \sum_p B_{p\, q}(t)\ c_p\, , \\
\cd_{p}& \  = \ \sum_q B_{p\, q}(t)\ \ctd_q\, , \qquad \qquad c_p \ = 
\ \sum_q B^*_{p\, q}(t)\ \ct_q\, ,
\end{align} 
Note that each transformation is given by a time-dependent $N\times N$
unitary matrix $B$. From the original group $U(2^N)/U(2)^N$ of dimensions 
$\sim 2^{2 N}$ we have restricted to the group $U(N)$ of dimensions $N^2$.
The components of a given $M$-particles state vector $v^\M_{p_1 p_2 \dots p_M}$ transforms under such a 
group according to 
\begin{equation} \label{model:vt}
\vt^\M_{q_1\dots q_M}(t) \ =\ \sum_{p_1, \dots ,p_M} B_{p_1\, q_1}(t)  B_{p_2\, q_2}(t) \dots 
B_{p_M\, q_M}(t)\, v^\M_{p_1 \dots p_M}(t).
\end{equation}

The strategy now is as follows. We analyse the entanglement of a vector state 
$v^\M_{p_1\dots p_M}$ with respect to the defining--TPS. Once 
we have an expression for the entanglement in terms of $v^\M_{p_1\dots p_M}$ we just 
have to 
substitute $v$ with $\vt$ as defined in \eqref{model:vt} to obtain the entanglement of the 
same state vector with respect to the new TPS defined by the Bogoliubov transformation $B$.

\subsection{Entanglement in the defining--TPS} \label{def}

Doing calculations
with the von Neumann entropy $S$ \eqref{lines:von} proves to be 
very cumbersom. Another measure of entanglement that will be used in this section
is the \emph{linear entropy}. For a system $A$ of density matrix $\rho_A$, this is
simply defined as 
\begin{equation} \label{model:linear}
S_{(A)}\ = \ 1 \, -\, {\rm Tr}_A (\rho_A^2)\, .
\end{equation}
Note that we indicate such a quantity with the same letter $S$ as the von Neumann entropy, except that the system to which it applies appears as a subscript inside round brackets. Linear entropy is still a good measure of entanglement for bipartite states, although one must take into account that it is ``rescaled'' with respect to the von Neumann entropy. The linear entropy is in fact bounded by
$0 \leq S_{(A)} \leq 1-1/D$ where $D$ is the dimension of space $\cH_A$ and 
the upper limit is reached, again, by maximally entangled states.

In the occupation number representation the density matrix is an object of $2N$ indeces;
for a pure state $|\psi\ket$ \eqref{model:psi}, for instance, it reads 
\begin{equation}
\rho_{\alpha_1\, \dots \, \alpha_N, \, \beta_1\, \dots\, \beta_N} = \psi^*_{\alpha_1\, \alpha_2\, \dots \, \, \alpha_N}\, \psi_{\beta_1\, \beta_2\, \dots \,  \beta_N}. 
\end{equation}
By starting from the defining--TPS \eqref{model:tps} we can choose a bipartition of $\cH$ 
into $2^L$ and $2^{N-L}$ dimensional systems and calculate the entaglement 
between these two parties. A partition is assigned by enumerating 
which of the $\mathbb{C}^2$ spaces is wanted on either sides; by definition,
in the partition 
\begin{equation} \label{model:partition}
P = \{l_1,\, \dots\, , \, l_L \} \qquad \text{with}\ L\leq N\, ,
\end{equation}
we are considering on the one side the $l_1^{\rm th}, l_2^{\rm th}, \dots , 
l_L^{\rm th}$ factors and, on the other side, all the other factors. 
The density matrix $\rho_P$ for the subsystem $P$ reads, in components,
\begin{equation}
({\rho_P})_{\al_{l_1}\dots \al_{l_L}, \ \be_{l_1}\dots \be_{l_L}} \
= \ (\rho)_{\al_1 \dots \al_{l_1} \dots \al_{l_L} \dots \al_N,\ \al_1 
\dots \be_{l_1} \dots \be_{l_L} \dots \al_N}\, ,
\end{equation}

The particle number representation is most useful to calculate entanglement, 
while the time evolution, as we are going to see in Sec. \ref{time},  is best studied on the ``fixed number of particles'' 
basis \eqref{model:mpar}. 
In order to go from the ``fixed number of particles'' basis to the 
occupation number representation we introduce a tensor 
$\Omega^\M_{\, p_1 p_2 \dots p_M} \in (\mathbb{C}^2)^{\otimes N}$ 
totally antisymmetric in the indeces $p_1 \dots p_M$
and of components either 0 or 1. Explicitly, for $p_1<p_2<\dots<p_N$, we define
$\Omega^\M$ as
\begin{equation}
(\Omega^\M_{\, p_1 p_2 \dots p_M})_\alphas \ \equiv\ 
\delta_{\al_1}^0 \, \delta_{\al_2}^0\, \dots \, \delta_{\al_{p_1}}^1\, 
\delta_{\al_{p_1+1}}^0 \, \dots 
\, \delta_{\al_{p_2}}^1 \, \dots \, \delta_{\al_N}^0\, .
\end{equation}
In the above equation
and in the following ones kronecker deltas have one subscript and a superscript for pure 
pictorial reasons.
A given state vector $v$ can be now written in the occupation number representation
as
\begin{equation} \label{model:vtopsi}
\psi_\alphas \ = \ \sum_{p_1 \dots p_M} v^\M_{p_1 \dots p_M}\,
(\Omega^\M_{\, p_1 \dots p_M})_\alphas \, .
\end{equation}
In order to calculate products of $\Omega$ tensors it is useful to introduce a sort of
generalized, antisymmetrized kronecker delta, defined as 
\begin{equation} \label{model:genkro}
\delta_{p_1 p_2 \dots p_M}^{q_1 q_2 \dots q_M} \ \equiv \ 
\delta_{p_1}^{q_1}\, \delta_{p_2}^{q_2}\, \dots \, \delta_{p_M}^{q_M}\, 
- \, \delta_{p_1}^{q_2}\, \delta_{p_2}^{q_1}\, \dots \, \delta_{p_M}^{q_M}\, 
+ \, \dots\ .
\end{equation}
We can then explicitly check the normalization condition \eqref{model:normpsi}:
\begin{align} \nonumber
\psi^*_{\alpha_1,\, \alpha_2,\, \dots \, ,\, \alpha_N}\, 
\psi_{\alpha_1,\, \alpha_2,\, \dots \, ,\, \alpha_N}\ =& 
 \sum_{\substack{
	p_1 \dots p_M\\
	q_1 \dots q_M}} v^{\M *}_{p_1 \dots p_M}\, v^\M_{q_1 \dots q_M}\, 
(\Omega^\M_{\, p_1 \dots p_M})_\alphas \, 
(\Omega^\M_{\, q_1 \dots q_M})_\betas \\ \label{model:norm}
=& \sum_{\substack{
	p_1 \dots p_M\\
	q_1 \dots q_M}} v^{\M *}_{p_1 \dots p_M}\, v^\M_{q_1 \dots q_M}\,  
\delta_{p_1 p_2 \dots p_M}^{q_1 q_2 \dots q_M}\\ \nonumber
=& \, M!\,  \sum_{p_1 \dots p_M} |v_{p_1 \dots p_M}|^2 \ =\ 1\ ,
\end{align}
where, in the last equality, \eqref{model:normv} has been used. In terms of 
the generalized kronecker delta \eqref{model:genkro} we can also express more general 
$\Omega$ products. With an increasing order of complexity we have:
\begin{equation}
(\Omega^{(2)}_{\, p_1 p_2})_{\al_1 \dots \al_l \dots \al_N}
(\Omega^{(2)}_{\, q_1 q_2})_{\al_1 \dots \be_l \dots \al_N}\ = \
\delta_{p_1 p_2}^{q_1 q_2} \, \times \, 
\begin{cases} 
\delta_{\al_l}^1 \delta_{\be_l}^1 \ \ \text{if} \ l\in \{p_1, p_2\} \\[2mm]
\delta_{\al_l}^0 \delta_{\be_l}^0 \ \ \text{if} \ l\notin \{p_1, p_2\}
\end{cases},
\end{equation}
\begin{equation}\label{model:product}
(\Omega^\M_{\, p_1 \dots p_M})_{\al_1 \dots \al_l \dots \al_N}
(\Omega^\M_{\, q_1 \dots q_M})_{\al_1 \dots \be_l \dots \al_N}\ = \
\delta_{p_1 \dots p_M}^{q_1 \dots q_M} \, \times \, 
\begin{cases} 
\delta_{\al_l}^1 \delta_{\be_l}^1 \ \ \text{if} \ l\in \{p_1, \dots , p_M\} \\[2mm]
\delta_{\al_l}^0 \delta_{\be_l}^0 \ \ \text{if} \ l\notin \{p_1, \dots , p_M\}
\end{cases}.
\end{equation}

Among all the possible partitions \eqref{model:partition} of the total Hilbert 
space we restrict from now on to the simple ones that single out just one -- 
the $l^{\rm th}$ -- $\mathbb{C}^2$ 
factor:
$P = \{ l\}$. We now want to calculate the corresponding reduced density matrix, $\rho^\M_l$,
for when the state vector of the Universe $v^\M$ belongs to a fixed number of particles 
subspace $\cH_M$. Such a quantity is a $2 \times 2$  matrix and, 
by equation \eqref{model:vtopsi}, has components
\begin{equation}
(\rho^\M_l)_{\al_l \be_l}\ =\  \sum_{\substack{
	p_1 \dots p_M\\
	q_1 \dots q_M}} v^{\M *}_{p_1 \dots p_M}\, v^\M_{q_1 \dots q_M}\, 
(\Omega^\M_{\, p_1 \dots p_M})_{\al_1 \dots \al_l \dots \al_N} \, 
(\Omega^\M_{\, q_1 \dots q_M})_{\al_1 \dots \be_l \dots \al_N}\, .
\end{equation}
Then, by using \eqref{model:product}, 
\begin{equation} \nonumber
(\rho^\M_l)_{\al_l \be_l}\ = \delta_{\al_l}^1 \delta_{\be_l}^1 \! \sum_{\substack{
	l \in \{p_1 \dots p_M\}\\
	\{q_1 \dots q_M\} }} v^{\M *}_{p_1 \dots p_M}\, v^\M_{q_1 \dots q_M}\, 
\delta_{p_1 \dots p_M}^{q_1 \dots q_M}
+ \delta_{\al_l}^0 \delta_{\be_l}^0 \! \sum_{\substack{
	l \notin \{p_1 \dots p_M\} \\
	\{q_1 \dots q_M\} }} v^{\M *}_{p_1 \dots p_M}\, v^\M_{q_1 \dots q_M}\, 
\delta_{p_1 \dots p_M}^{q_1 \dots q_M} ,
\end{equation}
where the subscripts of the sum symbols stay for ``sum only on the $M$-uples that contain  
$l$" and ``sum only on the $M$-uples that do not contain $l$" respectively. By 
\eqref{model:norm} the two sums add up to unity and we obtain therefore
\begin{equation}
(\rho^\M_l)_{\al_l \be_l}\ = \ 
\delta_{\al_l}^0 \delta_{\be_l}^0 + \left(\delta_{\al_l}^1 \delta_{\be_l}^1  -\delta_{\al_l}^0 \delta_{\be_l}^0 \right)\! \sum_{\substack{
	l \in \{p_1 \dots p_M\} \\
	\{q_1 \dots q_M\} }} v^{\M *}_{p_1 \dots p_M}\, v^\M_{q_1 \dots q_M}\, 
\delta_{p_1 \dots p_M}^{q_1 \dots q_M} 
\, .
\end{equation}
It is now straightforward to sum on each individual $M$-particles subspace and obtain the reduce density matrix for the general mixed state \eqref{model:block},
\begin{equation}
(\rho_l)_{\al_l \be_l}\ = \ 
\delta_{\al_l}^0 \delta_{\be_l}^0 + \left(\delta_{\al_l}^1 \delta_{\be_l}^1  -\delta_{\al_l}^0 \delta_{\be_l}^0 \right) \! \sum_{M = 0}^N\, p_M \sum_{\substack{
	l \in \{p_1 \dots p_M\} \\
	\{q_1 \dots q_M\} }} v^{\M *}_{p_1 \dots p_M}\, v^\M_{q_1 \dots q_M}\, 
\delta_{p_1 \dots p_M}^{q_1 \dots q_M} 
\, .
\end{equation}
By taking the square and then the trace of the above matrix we find the linear entropy
\eqref{model:linear} of the $l^{\rm th}$ $\mathbb{C}^2$ factor when the whole system is in the mixed state \eqref{model:block}. It reads
\begin{equation} \label{model:S}
S_{(l)} \ = \ 2 \left(\Si_l - \Si_l^2\right)\, ,
\end{equation}
where
\begin{equation} \label{model:Sigma}
\Si_l\ =\ \sum_{M = 0}^N\, p_M \sum_{\substack{
	l \in \{p_1 \dots p_M\} \\
	\{q_1 \dots q_M\} }} v^{\M *}_{p_1 \dots p_M}\, v^\M_{q_1 \dots q_M}\, 
\delta_{p_1 \dots p_M}^{q_1 \dots q_M} \, .
\end{equation}

\subsection{The time evolution} \label{model:time}

The time evolution of the state vectors is given by the Schroedinger equation
\begin{equation} \label{model:schroe}
\frac{d\, |\psi; t\ket}{d t} = - i H |\psi; t\ket \, .
\end{equation}
The time derivatives of the state vector's components $v^\M$ are obtained by directly 
substituting the expression for the state vector \eqref{model:mpar} and the explicit form
of the Hamiltonian \eqref{model:hamiltonian} into \eqref{model:schroe} and using the 
anticommutation relations \eqref{model:anti}.
We study only three cases: $(i)$ the time evolution of the one-particle states, $(ii)$ the 
time evolution of a generic state in the the free theory $(\g_{jklm} = 0)$ 
and $(iii)$ the time evolution of the two-particles states in the full theory.

$(i)$ The time evolution of the one-particle states $v_p$ is very simple since it is governed 
by the  free theory only. We obtain, in fact,
\begin{equation}
\dot{v}_p\ =\ -i\, \lambda_p\, v_p \, \quad \qquad \qquad 
v_p(t)\ =\ e^{-i\lambda_p t} v_p(0) .
\end{equation}

$(ii)$ If governed only by the free Hamiltonian $H_0$, the evolution of the generic 
$M$-particles state is also very simple, the evolutor operator being factorized into
$M$ pieces:
\begin{equation} \label{model:evolutor}
v^\M_{p_1 p_2 \dots p_M}(t) \ = \ e^{-i\lambda_{p_1} t} \, e^{-i\lambda_{p_2} t}\,  \dots 
\, e^{-i\lambda_{p_M} t} \, v^\M_{p_1 p_2 \dots p_M}(0) \, .
\end{equation}

$(iii)$ Of a two-particles vector $v_{p q}(t)$ we have only been able to 
to obtain the expressions of the derivatives in $t=0$ in terms of 
$v_{p q} \equiv v_{p q}(t=0)$. By making use also of the symmetries \eqref{model:sym1} 
and \eqref{model:sym2} we obtain
\begin{equation}\label{model:vdot}
\dot{v}_{pq}\ = \ -i\left[(\lambda_p+\lambda_q) v_{p q}\, - \, 2 \sum_{jk} \g_{pqjk}\, v_{jk}\right]\, ,
\end{equation}
\begin{equation} \label{model:vddot}
\ddot{v}_{pq}\ =\ -(\lambda_p+\lambda_q)^2 v_{p q}\, + \,  
2 \sum_{jk} (\lambda_p+\lambda_q + \lambda_j +\lambda_k)\, \g_{pqjk}\, v_{jk}\,
-\, 4 \sum_{jkj'k'} \g_{pqjk} \ \g_{jkj'k'}\ v_{j'k'}\, .
\end{equation}

\subsection{Entanglement evolution} \label{model:evolution}

By substituting $\vt$ (defined in equation \eqref{model:vt}) to $v$ in equations \eqref{model:S} and \eqref{model:Sigma} we find the entanglement entropy of the subsystem $l$ in the new TPS (defined by the Bogoliubov transformation $B_{pq}(t)$) when the total system is in the state
$v(t)$.

$(i)$ Consider first the one-particle case. The transformation \eqref{model:vt} reduces to
\begin{equation}
\vt_q(t)\ =\, \sum_p B_{p q}(t)\, v_p(t)\, ,
\end{equation}
and the quantity $\Si_l$ defined in \eqref{model:Sigma} reads
\begin{equation} \label{model:Sig2}
\Si_l(t)\ =\ \sum_{pq} B^*_{pl}(t) B_{q l}(t) e^{-i(\lambda_q - \lambda_p) t}
v^*_p(0) v_q(0)\, .
\end{equation}
In order to test conjecture (V) (at the end of Sec. \ref{lines:local}) we have to look 
for the TPS that minimizes the creation of entanglement. This is trivially accomplished by
dividing $B_{p q}(t)$ in a time independent part $B_{p q}(0)$ and a time dependent part and 
taking the latter to be a diagonal matrix with $e^{i \lambda_q t}$ on the diagonal:
\begin{equation}\label{model:int}
B_{ql}(t) \ = \ B_{ql}(0)\, e^{i \lambda_q t}\, .
\end{equation}
We call a TPS of this kind, no matter what the time-independent part $B_{ql}(0)$ is, an
``\emph{interaction}--TPS'', since it reminds the interaction picture of quantum mechanics
where the ``free'' part of the evolution is reabsorbed in the definition of the 
state vectors. A particular interaction--TPS is defined through \eqref{model:int} by 
a particular choice of the time-independent $U(N)$ matrix $B_{ql}(0)$ that from 
now on will be simply indicated as $B_{ql}$.

By substituting \eqref{model:int} into \eqref{model:Sig2} 
the time dependence is totally reabsorbed and the locality conjecture (V) is satisfied in a trivial way: 
\emph{in every interaction--TPS the totally factorized one-particle states  remain 
disentangled during the entire evolution}. According to the general ideas of Sec.
\ref{lines}, therefore, there are \emph{no coincidence relations between localized systems} in this case and 
a spacetime cannot be defined.

$(ii)$ The M-particles free case, because of the factorized form of the evolutor 
\eqref{model:evolutor} is very similar to the one just considered. The time evolution can, 
again, be reabsorbed by any interaction-TPS and therefore no spacetime can be 
operationally defined in 
this case. This is a nice null result: only interacting theories can operationally
define a spacetime!

$(iii)$ Let's now consider
two-particles states evolving with the complete Hamiltonian
\eqref{model:hamiltonian}. In this case the evolutor is not factorizable and therefore
no TPS can totally reabsorb the effects of time evolution. The quantity $\Si_l$ 
\eqref{model:Sigma} reads, in general,
\begin{equation} \label{model:Sig3}
\Si_l(t)\ =\ 4 \sum_p|\vt_{pl}(t)|^2 \ = \ 4 \sum_{p\, q_1 \dots q_4}
B^*_{{q_1} p}(t) B^*_{{q_2} l}(t) B_{{q_3} p}(t)B_{{q_4} l}(t) 
v^*_{{q_1}{q_2}}(t)v_{{q_3}{q_4}}(t)\, .
\end{equation}
It is reasonable to fix again the time dependence of the TPS
according to ansatz \eqref{model:int}. In fact, in the perspective of minimizing the 
creation of entanglement in
general statistical mixtures \eqref{model:block} of different particles contents, it looks
reasonable to restrict to interaction--TPSs \eqref{model:int}
which already minimize entanglement at least in the one-particle sector.
Whether or not this ansatz is a sensible
choice also for the two-particles subspace alone is going to be checked \emph{a posteriori}.

In order to use conjecture (V) we have to prepare an initial state vector 
totally factorized with respect 
to some interaction--TPS. Apart from an irrelevant phase, 
the possible choices amount to all possible pairs of sites 
$\{ \mun , \md \}$ that we want to be initially occupied:
\begin{equation} \label{model:initial}
\vt_{pq}(0)\ = \ \delta_{[p}^\mun \, \delta_{q]}^\md \ \equiv \ 
\frac{1}{2}\left(\delta_{p}^\mun \, \delta_{q}^\md - \delta_{q}^\mun \, \delta_{p}^\md \right).
\end{equation}
Note that, from now on, all the quantities that we are going to calculate depend on three 
indeces: $l$ is the party with respect to which we calculate the entanglement, while $m_1$ and 
$m_2$ are the parties that characterize the initial state \emph{i.e.} the two sites
``occupied'' in the initial configuration of our toy-Universe.

The tendency to entanglement between the party $l$ and the rest of the Universe
is measured by the time derivatives of $S_{(l)}$ at $t=0$. From \eqref{model:S} we get
\begin{equation} \label{model:sder}
\dot{S}_{(l)}\ =\ 2 \dot{\Si}_l \left(1 - 2 \Si_l \right), \qquad 
\ddot{S}_{(l)}\ =\ 2 \ddot{\Si}_l \left(1 - 2 \Si_l \right) - 4 \dot{\, \Si}^2_l\, .
\end{equation}
By substituting \eqref{model:initial} into \eqref{model:Sig3} we obtain the value of $\Si_l$ at $t=0$,
\begin{equation}
\Si_l(0)\ =\ \delta_l^\mun \, + \, \delta_l^\md \, .
\end{equation}
For an interaction--TPS \eqref{model:int}, moreover,
\begin{equation} \label{model:bder}
\dot{B}_{q l}(t)|_{t = 0} \ = \ i \lambda_q  B_{q l}\, , \qquad
\ddot{B}_{q l}(t)|_{t = 0} \ = \ - \lambda_q^2  B_{q l}\, .
\end{equation}
One can now take the derivative of \eqref{model:Sig3} and express the time 
derivatives of $v$ and $B$ with the aid of \eqref{model:vdot} 
and \eqref{model:bder} respectively. With the help of the symmetries \eqref{model:sym1} and \eqref{model:sym2}
of the model one can show that, for the initial state \eqref{model:initial},
$\dot{\Si}_l(0)\ = \ 0$, both when the party $l$ is one of the two initially 
occupied sites, $l\in \{ \mun,\md \}$ and when it's not, $l\notin \{ \mun,\md \}$.
It trivially follows from \eqref{model:sder}  that also $\dot{S}_{(l)}(0) = 0$, and that 
\begin{equation}
\ddot{S}_{(l)} (0) \ =\ \begin{cases}
	- 2 \ddot{\Si_l}(0) \ \ \text{if} \ l\in\{ m_1, m_2\}\\[2mm]
	\ \ 2 \ddot{\Si_l}(0) \ \ \text{if} \ l\notin\{ m_1, m_2\}
		\end{cases}\, .
\end{equation}
Since the above \emph{second derivative} is generally non-null, we use this 
quantity to measure the ``tendency to entanglement'' of a given TPS.
As before, to calculate $\ddot{S}_{(l)}(0)$,  one can derive expression \eqref{model:Sig3} twice
and use \eqref{model:vdot} 
and \eqref{model:bder}.
As a warming up, it is useful to calculate $\ddot{\Si_l}(0)$ in the  case
of the defining--interaction--TPS \emph{i.e.} when $B_{ql} = \delta_q^l$. We obtain:
\begin{equation} \label{model:defining}
\ddot{\Si_l}(0)\ =\ 16\left[2\sum_p|\g_{\mun \md p l}|^2 - (\delta_l^\mun + \delta_l^\md)
\sum_{jk} |\g_{\mun \md j k}|^2
\right]\, .
\end{equation}
The same expression in a general interaction-TPS case is also straightforward but 
a bit lengthier; it is worth divide it into the two pieces: 
\begin{equation} \label{model:sigddot}
\ddot{\Si_l}(0) \ = \ 8\, A
+ 4 \, B \, , 
\end{equation}
where
\begin{equation} \label{model:A}
A\ = \ \sum_p \dot{\vt}_{pl}^*\dot{\vt}_{pl}\ = \ 4 \sum_{\substack{
	\text{all but}\\
	l\, m_1\, m_2}} \g_{p_1 p_2 j_1 j_2}\, \g^*_{p_1 q k_1 k_2}\,
B_{p_2 l}\, B^*_{q l}\, B^*_{j_1 [m_1} B^*_{j_2 m_2]}\, B_{k_1 [m_1} B_{k_2 m_2]}
\end{equation}
\begin{multline} \label{model:B}
B\ =\ \sum_p (\vt^*_{pl} \ddot{\vt}_{pl} + {\rm c. c.})\ = \\
-2 \!\!\sum_{\substack{
	\text{all but}\\
	l\, m_1\, m_2}}\! \g_{p_1 p_2 k_1 k_2}\, \g_{k_1 k_2 j_1 j_2}
\left[(\delta_l^\md B_{p_1 m_1} B_{p_2 m_2} - \delta_l^\mun B_{p_1 m_2} B_{p_2 m_1})
B^*_{j_1 [m_1} B^*_{j_2 m_2]}\, + \, {\rm c.c.}\right]\, .
\end{multline}

\subsection{An example: the one dimensional Heisenberg spin chain} \label{heis}

The above expressions, obtained also with the aid of the symmetries 
\eqref{model:sym1} and \eqref{model:sym2}, do not depend on the free part of the Hamiltonian 
but only on the form of the interaction kernel $\g_{j k l m}$. If this pre-geometric approach
is correct the tensorial structure of $\g_{j k l m}$ should therefore encode the very dimensionality 
of spacetime since, as a matter of fact,
no reference to the spacetime dimensions has been made so far\footnote{Interestingly, 
delta-like momentum-conserving kernels and their very three dimensional
structures have been argued \cite{frog} to originate from general fermionic and bosonic theories 
in some appropriate low-energy limit.}. 
In this section we consider the example of a 
one-dimensional XXY Heisenberg spin chain which can be proved \cite{frad}
to be equivalent to the fermionic model \eqref{model:hamiltonian} for 
the following choices of the kernels:
\begin{equation}
\lambda_j \ = \ \cos\left(\frac{2 \pi j}{N}\right)\, ,
\end{equation}
\begin{equation} \label{model:gamma}
\g_{jklm} = \Delta\,  \delta(j+k-l-m)\left[\cos \frac{2 \pi (j-l)}{N}
-\cos\frac{2 \pi (k-l)}{N} \right]\, ,
\end{equation}
where $\Delta$ is just a positive coupling parameter.
In this case the following useful relation holds:
\begin{equation}
\sum_{k_1 k_2} \g_{j l k_1 k_2}\, \g_{k_1 k_2 m n} \ = \
\Delta\, N\, \g_{jlmn}\, .
\end{equation}

We called ``defining--TPS'' the one associated to the operators $c_j$ and $\cd_j$, with respect 
to which the free Hamiltonian $H_0$ has the diagonal form \eqref{model:hamiltonian};
with respect to a generic Bogoliubov transformed set of operators $\ct_j$ and $\ctd_j$, 
in fact, $H_0$ is generally expressed as a non-diagonal quadratic form. 
The set of operators $c_j$ that render $H_0$ diagonal are usually associated with the 
``momentum basis'' \emph{i.e.} are the ones that create or destroy a quantum with 
a definite value $j$ of the momentum.
According to our conjecture (V) the localized subsystems should be singled out by the 
interaction--TPS $B_{pq}$ that minimizes $\ddot{S}_{(l)}(0)$.
Such a minimization problem is not elementary
since the space to search for the minimum 
is that of the $N \times N$ unitary matrices $U(N)$, and trying to solve it by direct 
analytic methods proves to be very cumbersome.
Here we show that $\ddot{S}_{(l)}(0)$ is null in the 
interaction--TPS that, in this Heisenberg model, is usually associated with the ``position basis'', 
\emph{i.e.} the one in which $B_{pq}$ is just a Fourier transform. As a consequence, 
since the linear entropy is a positive 
definite quantity, such a TPS is in fact most probably a minimum, although it is not clear 
whether degenerate or not.
We  have checked, on the other hand, that a null second derivative of $S$ is neither a trivial 
result nor a general occurrence; for instance, in the defining--interaction--TPS, by 
substituting \eqref{model:gamma} into \eqref{model:defining} we find
\begin{equation} \nonumber
\ddot{\Si_l}(0) = 16 \Delta^2 \left[2 \left(\cos \frac{2 \pi (m_2-l)}{N}
-\cos\frac{2 \pi (m_1-l)}{N} \right) - N(\delta_l^{m_1}+\delta_l^{m_2}) 
\left(1 - \cos \frac{2 \pi (m_1 - m_2)}{N}\right)\right]\, .
\end{equation}

In order to calculate \eqref{model:sigddot} for the interaction--TPS associated with the 
position basis, we first note that the  
coefficient $A$ in eq. \eqref{model:A} is better written as
\begin{equation}
A\ = \ \sum_q {\cal C}_{l m_1 m_2 q}\, {\cal C}^*_{l m_1 m_2 q},
\end{equation}
where
\begin{equation} 
{\cal C}_{l m_1 m_2 p}\ \equiv \ 2 \sum_{q j_1 j_2} \g_{p q j_1 j_2} \,  B^*_{j_1 [m_1} 
B^*_{j_2 m_2]} B_{ql} .
\end{equation}
By substituting the Fourier transform 
\begin{equation} \label{model:fourier}
B_{q l}\ =\ N^{-1/2} e^{-2 \pi i ql/N}
\end{equation}
and expression \eqref{model:gamma} for the kernel $\g$ we find
\begin{multline}
{\cal C}_{l m_1 m_2 p}\ = \\ \Delta N^{1/2} \left[e^{2 \pi i p m_1 /N} 
(\delta_l^{m_1 -1} \delta_l^{m_2} + \delta_l^{m_1 +1} \delta_l^{m_2})\, -
\, e^{2 \pi i p m_2 /N} (\delta_l^{m_1} \delta_l^{m_2-1} + \delta_l^{m_1} \delta_l^{m_2 +1})
\right]\, ,
\end{multline}
from which
\begin{equation}
A \ = \ \Delta^2 N^2 (\delta_{m_1}^{m_2+1} + \delta_{m_1}^{m_2 -1}) 
(\delta_l^{m_1} + \delta_l^{m_2}) .
\end{equation}

On the other hand, by substituting \eqref{model:fourier} and 
\eqref{model:gamma} into \eqref{model:B} one obtains, after some simplifications,
\begin{equation}
B\ =\ - 2 \Delta^2 N^2 (\delta_{m_1}^{m_2+1} + \delta_{m_1}^{m_2 -1}) 
(\delta_l^{m_1} + \delta_l^{m_2}),
\end{equation}
from which
\begin{equation}
\ddot{S}_{(l)}(0) \ = \ \pm 2 \ddot{\Si_l}(0)\ =\ 0 \, .
\end{equation}

\section{ABOUT GRAVITY} \label{last}

We have argued that the usual description of physical processes as ``taking place" in a spacetime can be entirely derived on the basis of the informational content of quantum theories formulated without any allusion to a pre-existing spacetime.
The general ideas of sections \ref{relational} and \ref{lines} rely only on the structure of product Hilbert spaces and, therefore, apply in principle to any quantum theory (quantum field theory, string theory, etc \dots). Of the emerging spacetime picture we have explicitly drawn only $(i)$ the class (or classes) of \emph{localized systems} (Sec. \ref{lines:local}) and
$(ii)$ the \emph{coincidence relations} (Sec. \ref{cont}) that such systems undergo
between each other; any other metric relation should be
derivable from this premise (see
the remarks at the beginning of Sec. \ref{cont} and at the end of Sec. \ref{time})
although such an explicit derivation is beyond the purposes of the present work.

By applying the locality conjecture (V) of Sec. \ref{lines:local} to the elementary 
case of a one-dimensional Heisenberg spin chain, in Sec. \ref{heis} we recover  as 
\emph{local} those degrees of freedom which are normally associated with ``position''. 
The only reference to the dimensionality of the model is in the structure
choosen for the interaction kernel $\g_{jklm}$; the spatial picture 
of a one-dimensional chain of ``localized systems'' is an output of the 
proposed constructive procedure.
This is an encouraging result that we plan to extend to more realistic (\emph{e.g.}
three-dimensional) theories.

We note also that anytime we try to apply 
conjecture (V) something interesting happens. 
The corresponding minimization problem  is not unique and, in fact,
there are as many minimization problems -- and as many solutions -- as the possible initial states $\vt^\M(0)$ that can be choosen for the system-Universe. In Sec. \ref{model:evolution} we explicitely calculated the tendency of entanglement $\ddot{S}_{(l)}$
only for the states belonging to the one-particle $(M=1)$ and two-particles $(M=2)$ subspaces.
Some relevant differences already show up between these two types of configurations: for one-particle states
the minimization problem is trivially satisfied by all types of \emph{interaction}--TPSs 
(see eq. \ref{model:int} and discussion thereafter), while for two-particles states the minimum -- even if degenerate -- is surely much more constrained. 
More generally, the class of systems that we define as \emph{localized} -- and thus the emerging spacetime picture itself -- depends on the choosen initial state $\vt^\M(0)$ and, therefore, on the \emph{matter content} of the system-Universe.

Whether or not gravity itself is contained in this mechanism (\emph{i.e.} the emergent spacetime has already the metric properties prescribed by GR for that given matter content) cannot be, at this stage, much more than a speculation, although, admittedly, exploring and checking this possibility is the original and wider aim of this approach. It is worth noting that the idea of gravity as an emergent phenomenon by-produced by 
the other quantum fields is not at all new.
In the usual framework of quantum field theory,
gravitational interaction terms of the cut-off size 
are \emph{induced} \cite{sak,adler} by radiative corrections, although the very same 
mechanism generally produces also a huge -- and unobserved -- comological constant term 
\cite{wei}. In the more subtle approach of \cite{vene}, where no reference is made 
to a metric field 
and fundamental fermions to define a connection are used instead, the
gravitational theory induced at low energies is immune to the cosmological constant problem.
More generally, it is arguable that a thin or null vacuum energy is not necessarily a problem 
if the metric degrees of freedom are composite or ``collective'' instead of fundamental. 
Analogies \cite{carlos} with the physics of condensed matter systems have been used to 
support 
this view \cite{volo}: some fields, such as the sound waves in a crystal or in a superfluid, 
are just collective modes of more fundamental and quantum mechanical degrees of freedom.
The quantum properties of the fundamental components -- vacuum energy included -- 
are already embodied \cite{volo2} in the definition of such collective modes and, therefore, 
do not affect their dynamics.

\section*{ACKNOWLEDGMENTS}
I'm grateful to Sergio Cacciatori for very useful discussions. I also thank Bruno Carazza, Roberto Casero, Thibault Damour, Roy Maartens, Roberto De Pietri, Carlo Rovelli, Michele Vallisneri, David Wands, Alberto Zaffaroni and Paolo Zanardi.
This work started at the Physics Department of the University of Milano Bicocca and has
been supported, at a later stage, by a Marie Curie Fellowship under contract number
MEIF-CT-2004-502356.


\end{document}